\documentclass[letterpaper,aps,prd,preprint,showpacs,nofootinbib,
superscriptaddress]{revtex4}
\pdfoutput=1
\usepackage{graphics}
\usepackage{graphicx}
\usepackage{amssymb}
\usepackage{color}

\newcommand{\nc}{\newcommand}

\nc{\postscript}[2]{\setlength{\epsfxsize}{#2\hsize}\centerline{\epsfbox{#1}}}
\nc{\beq}{\begin{equation}}   \nc{\eeq}{\end{equation}}
\nc{\bea}{\begin{eqnarray}}   \nc{\eea}{\end{eqnarray}}
\nc{\baa}{\begin{array}}      \nc{\eaa}{\end{array}}
\nc{\bit}{\begin{itemize}}    \nc{\eit}{\end{itemize}}
\nc{\ben}{\begin{enumerate}}  \nc{\een}{\end{enumerate}}
\nc{\bce}{\begin{center}}     \nc{\ece}{\end{center}}
\nc{\non}{\nonumber}

\begin{document}

\begin{flushright}

 \mbox{\normalsize \rm CUMQ/HEP 165}\\
        \end{flushright}
\vskip 20pt

\title {\bf  Radion Phenomenology with 3 and 4 Generations}

\vskip 20pt

%\author{AAA?}
\author{Mariana Frank\footnote{mfrank@alcor.concordia.ca}}
\affiliation{Department of Physics, Concordia University\\
7141 Sherbrooke St. West, Montreal\\ Quebec, CANADA H4B 1R6
}
\author{
Beste Korutlu\footnote{bkorutlu@physics.concordia.ca}}
\affiliation{Department of Physics, Concordia University\\
7141 Sherbrooke St. West, Montreal\\ Quebec, CANADA H4B 1R6
}
\author{Manuel Toharia\footnote{mtoharia@physics.concordia.ca}}
\affiliation{Department of Physics, Concordia University\\
7141 Sherbrooke St. West, Montreal\\ Quebec, CANADA H4B 1R6
}

\date{\today}

\begin{abstract}
We study radion phenomenology in an warped extra-dimension scenario with
Standard Model fields in the bulk, with and without an additional fourth
family of
fermions. The radion couplings with the fermions  will be generically
misaligned with respect to the Standard Model fermion mass matrices, therefore
producing some amount of flavor violating couplings and potentially
influencing production and decay rates of the radion. 
Simple analytic expressions for the radion-fermion couplings are obtained with
three or four families. We also update and analyze the current
experimental limits
on radion couplings and on the model parameters, again with both three and four
families scenarios. We finally present the modified decay branching ratios of
the
radion with an emphasis on the new channels involving flavor diagonal and flavor
violating decays into fourth generation quarks and leptons.

\end{abstract}


\maketitle

%%%%%%%%%%%%%%%%%%%%%%%%%%%%%%%%%%%%%%%%%%%%%%%%%%%%%%%%%%%%%%%%%%%%%%%%%%%%%%%%
%%%%%%%%%%%%%%%%%%%%%%%%%%%%%%%%%%%%%%%%%%%%%%%%%%%%%%%%%%%%%%%%%%%%%%%%%%%%%%%%
\section{Introduction}

Warped extra dimensional models have been introduced to explain  the
origin of the  discrepancy between  Planck scale and the electroweak scale
\cite {RS}. In the original scenario, two branes are introduced, one with an
energy scale set at the Planck scale, the other at the TeV scale, and with
the Standard Model (SM) fields localized  on the TeV brane and gravity
allowed to propagate in the bulk. However, in this 
scenario, higher dimensional operators of the IR fields in the 4-dimensional
effective theory are only suppressed by TeV scales, leading to large flavor
violation and rapid proton decay. 

Allowing SM fermions and gauge fields to
propagate in the bulk effectively suppresses such operators and can
also be used to explain the fermion mass hierarchy by fermion
localization \cite{a, bulkSM}. 
The drawback is that, in minimal models, excitations of the bulk
fields are subjected to tight bounds from precision electroweak tests
\cite{EWPTmodel} and from flavor physics \cite{Weiler}, and
constrained to be heavier than a few TeV, making it very hard to
produce and observe heavy resonances of these masses at the LHC.
 
One hope to observe new states from these scenarios might be the
radion graviscalar and its associated phenomenology. The radion is a
scalar field associated with fluctuations in the size of the extra
dimension, and is a novel feature of warped extra dimensional
models \cite{RS}. The mass of the radion depends on the mechanism that
stabilizes the size of the extra dimension. In a simple model with a
bulk scalar which generates a vacuum expectation value (VEV), the
radion field emerges as a pseudo-Goldstone boson associated with
breaking of translation symmetry
\cite{Goldberger:1999uk}. Generically, the radion may be the lightest
new state in an RS-type setup, with its mass suppressed with respect
to KK fields by a volume factor of $\sim 40$, at least in the small
backreaction limit \cite{CGK}. This might put its mass between a few tens
to hundreds of GeV, with couplings allowing it to have escaped 
detection at LEP, and consistent with precision EW data
{CHL}. 
In general, the radion couplings are similar to Higgs couplings in
that they are proportional to the mass of the particles it couples
with. Moreover the radion field can mix with the Higgs boson after
EWSB, which involves another parameter, the coefficient of the
curvature-scalar term \cite{GRW}. 
Radion phenomenology with and without Higgs-radion mixing has been
discussed in several papers \cite{history,CHL}. More recently it has
been shown that a tree-level  misalignment between the flavor
structure of the Yukawa couplings of the radion and the fermion mass
matrix will appear when the fermion bulk parameters are not all
degenerate \cite{Azatov:2008vm}. The mechanism responsible for these
flavor-changing neutral currents (FCNC's) is different than the one
producing Higgs mediated FCNC's in these same models
\cite{Agashe:2009di,Azatov:2009na}. 

New data seems to indicate some inconsistencies with  the SM predictions as
pertaining to the  the third generation    \cite{newcdf}. The simplest
explanation is that  the Cabibbo-Kobayashi-Maskawa (CKM) mixing matrix deviates
from the standard, three-generation form \cite{Lunghi:2010gv}.   A simple
extension of the SM to four generations, SM4,  (adding a new family of quarks
and leptons, mirroring the existing ones) alleviates the problem, and may offer
resolutions to  some other outstanding problems in the SM \cite{holdom}. 

Recently, we have shown that, if the fourth generation is incorporated
into warped space models, the flavor-changing couplings of the Higgs
boson can be enhanced, and both the production and decays of the Higgs
bosons and the decay pattern of the heavy quarks and leptons is
altered significantly with respect to the patterns expected in SM4,
thus giving rise to distinguishing signals at the colliders
\cite{Frank:2011rw}. It is thus expected that in a warped scenario
with extra generations (seen as a natural extension of the warped
space model), the flavor-changing couplings of the radion will also
yield characteristic signals at colliders.

There are several reasons to perform a separate study for the radion
flavor-changing interactions. A priori, we expect the radion phenomenology to be
very different from the Higgs boson in four generations. As will be
explained later in the text, the production cross section for the Higgs
bosons in 4 generations is enhanced by a factor of about 10, while the
production cross section for the radion will remain essentially
unchanged by the presence of an extra family. Also, contrary to the
Higgs phenomenology in these models \cite{Frank:2011rw}, exotic flavor violating
decays of heavy quarks into radions $Q\to \phi q\ $ should be highly
suppressed with the new flavor violating couplings of the
radion. These will become important in radion decays into quarks $\phi
\to q q, \,qq'$ as well as into leptons $\phi \to \tau^\prime \tau$
and $\phi \to  \nu_{4} \nu_{\tau}$. 
Very recent data from ATLAS \cite{ATLASHnew} and CMS \cite{CMSHnew}
experiments at the LHC indicate that a Higgs boson in a scenario with
four generations must be very heavy. These measurements do not
affect the radion mass directly, but will set limits on the combined
radion mass--interaction scale parameter space. 
While we stated that the phenomenology with
three and four generations is quite similar, there are (new)  FCNC
effects of fourth generation quarks and leptons interacting with the
radion. 

Our paper is organized as follows. In the next section (Sec.
\ref{sec:themodel}), we
briefly review the warped model with fermions in the bulk,
concentrating in particular on the misalignment between the fermion
masses and radion-fermion Yukawa couplings. We describe in more detail
the flavor structure with four families in Sec \ref{sec:4generations},
and proceed to evaluate the radion FCNC couplings, presenting both
analytical and numerical results in Sec. \ref{sec:FCNC}. We discuss
phenomenological constraints in Sec. \ref{sec:pheno} and branching
ratios in Sec. \ref{sec:decays}. We summarize and conclude in
Sec. \ref{sec:conclusion}. Some details of our calculations are left to
the Appendix.

%%%%%%%%%%%%%%%%%%%%%%%%%%%%%%%%%%%%%%%%%%%%%%%%%%%%%%%%%%%%%%%%%%%%%%%%%%%%%%%%
%%%%%%%%%%%%%%%%%%%%%%%%%%%%%%%%%%%%%%%%%%%%%%%%%%%%%%%%%%%%%%%%%%%%%%%%%%%%%%%%
\section{The model}
\label{sec:themodel}

The radion graviscalar can be thought of as a scalar component of the
5D gravitational perturbations, and basically it tracks fluctuations
of size of the extra-dimension (i.e. its ``radius'').
The AdS metric including the scalar perturbation
$F$ corresponding to the effect of the radion is given in the RS
coordinate system by \cite{Charmousis}
 \begin{equation}
ds^2 = e^{-2 (A + F)} \eta_{\mu\nu} - (1+2F)^2 dy^2
= \left(\frac{R}{z}\right)^2\left( e^{-2F} \eta_{\mu\nu} dx^\mu dx^\nu  -
(1+2F)^2 dz^2
\right),
\end{equation}
where $A(y) =k\,y$. Note that the perturbed metric is no longer
conformally flat, even in $z$ coordinates.
At linear order in the fluctuation, $F$, the
metric perturbation is given by
\begin{equation}
\delta( ds^2 ) \approx -2 F \left( e^{-2A}
\eta_{\mu\nu} dx^\mu dx^\nu + 2~dy^2 \right) = -2 F \left( \frac{R}{z}
\right)^2 \left( \eta_{\mu\nu}dx^\mu dx^\nu + 2 dz^2\right).
\label{radionpert}
\end{equation}
In the absence of a stabilizing mechanism, the radion is precisely
massless, however it was shown that the addition of a bulk scalar
field with a vacuum expectation value (VEV) leads to an effective
potential for the radion after taking into account the back-reaction
of the geometry due to the scalar field VEV profile~\cite{CGK}.  In
the analysis that follows, we assume that this back-reaction is small,
and does not have a large effect on the 5D profile of the radion.

The relation between the
canonically normalized 4D radion field $\phi(x)$ and the metric
perturbation $F(z,x)$ is given by
\begin{equation}
F(z,x) = \frac{1}{\sqrt{6}} \frac{R^2}{R'} \left( \frac{z}{R}
\right)^2 \phi(x) =
\frac{\phi(x)}{\Lambda_\phi}\,\left(\frac{z}{R'}\right)^2,
\end{equation}
with $
\Lambda_\phi \equiv \frac{\sqrt{6}}{R'}$ the radion interaction scale.

The 5D interaction terms between the radion and the bulk SM fermion
fields are given by the action: 
\begin{eqnarray}
\label{fermionaction}
&&\hspace{-.5cm} S_{\text{fermion}}\!=\!\int d^4x dz \sqrt{g} \Big[
\frac{i}{ 2} \left({ \bar{\cal Q}} _i \Gamma^A {\cal D}_A {\cal Q}_i -
{\cal D}_A {\bar{ \cal Q}}_i \Gamma^A {\cal Q}_i\right)  \nonumber \\
&&\hspace{-.5cm} + \frac {c_{q_i}} { R} {\bar{ \cal Q}}_i {\cal Q}_i 
- \frac {c_{u_i}} { R} {\bar{ \cal U}}_i {\cal U}_i - \frac {c_{d_i}} { R}
{\bar{ \cal D}}_i {\cal D}_i  +\left(Y_{ij}\sqrt{R}\ {\bar{\cal Q}}_i
{\cal H} {\cal U}_j + h.c.\right)
\Big],
\end{eqnarray}
where $\displaystyle \frac{c_{q_i}}{R}$,  $\displaystyle \frac{c_{u_i}}{R}$ and
$\displaystyle \frac{c_{d_i}}{R}$ are the 5D fermion
masses, and we choose to work in the basis where these are diagonal in 5D
flavor space. The Higgs boson in the bulk acquires a VEV $v(z)$ localized
towards the IR 
brane thus solving the Planck-weak hierarchy problem. 

One can express the 5D fermions in two component notation, 
\begin{eqnarray}
{\cal Q}_i=\left( \begin{array}{c}{\cal Q}^i_L\\ {\bar{\cal Q}}^i_R
\end{array} \right)~,~~~ {\cal U}_i=\left(\begin{array}{c}{\cal U}^i_L\\
{\bar{\cal U}}^i_R \end{array}\right)~,~~~
{\cal D}_i=\left(\begin{array}{c}{\cal D}^i_L\\ {\bar{\cal D}}^i_R
\end{array}\right)~,
\end{eqnarray}
and perform a ``mixed'' KK decomposition as
\begin{eqnarray}
{\cal Q}^i_{L}(x,z) & =&  \sum_j Q_{L}^{ij}(z)\, q_L^j (x), \qquad {\bar{\cal
Q}}^i_R(x,z) =  \sum_j Q_{R}^{ij}(z)\, \bar{u}_R^j (x),\nonumber \\
{\cal U}^i_L(x,z)& =& \sum_j U_{L}^{ij}(z) q^j_L(x), \qquad \bar{\cal
U}^i_R(x,z)= \sum_j U_{R}^{ij}(z) \bar{u}^j_R(x), \nonumber \\
{\cal D}^i_L(x,z)& =& \sum_j D_{L}^{ij}(z) q^j_L(x),\qquad \bar{\cal
D}^i_R(x,z)= \sum_j D_{R}^{ij}(z) \bar{d}^j_R(x).
\label{kkdecomp}
\end{eqnarray}
Here  $q^j_L(x),\ u^j_R(x)$ and $d^j_R(x)$ are the 4D fermions
and  $Q^{ij}_{L,R}(z), \ U^{ij}_{L,R}(z)$ and $D^{ij}_{L,R}(z)$ are the
corresponding profiles along the extra
dimension. The fields $q_L^i(x), u_R^j(x)$ and $d_R^j(x)$ satisfy the
Dirac equation
\begin{eqnarray} 
-i \bar{\sigma}^{\mu} \partial_\mu q_L^{i} + m^u_{ij}\, \bar{u}_R^j &=& 0, \\
-i \sigma^{\mu} \partial_\mu {q}_L^i + m^d_{ij}\, \bar{d}_R^j& =& 0.
\end{eqnarray}
The 4D SM fermion mass matrix $m_{ij}$ is the eigenvalue which emerges
from the solution of the coupled bulk equations of motion, and is not
necessarily 
diagonal in flavor space. 
The couplings between the radion and SM fermions can be obtained by
inserting the perturbed metric  and the 5D fermion
KK decompositions of Eq.~(\ref{kkdecomp}) into the action
of Eq.~(\ref{fermionaction}).
We proceed by using a perturbative approach in treating the 4D
fermion masses $m_{ij}$ as small expansion parameters and keeping
only first order terms.
A 5D bulk Higgs is considered and its field perturbation contains
itself some radion degree of freedom.
Including all the contributions, the radion coupling to fermions can
be expressed finally as  
\begin{eqnarray}
\label{Radioncoupling}
\hspace{-2cm}&&-\frac{\phi(x)}{\Lambda_\phi} \left(q_L^{i} u_R^{j} +
\bar{q}_L^{i} \bar{u}_R^{j} \right)  {m^u_{ij}}
\left[{\cal I}({c_{q_i}}) + {\cal I}({c_{u_j}})\right] +( u \rightarrow d), 
\end{eqnarray}
with the definition
\begin{eqnarray}
\label{defI}
{\cal I}(c)= \left[\frac{(\frac{1}{2}-c)}{1-{(R/R')}^{1-2c}}+c\right]
\approx \Big\{ \begin{array}{c} c
  \, \,(\, c\, > \,1/2\,) \\ \frac{1}{2}\,\, (\, c \,<
  \,1/2\,) \end{array}.
\end{eqnarray}
This result from \cite{Azatov:2008vm} is consistent with the original
calculation obtained for the case of a brane Higgs and a single family
of fermions in \cite{CHL}. 

Non-universalities in the term $\left[{\cal I}({c_{q_i}}) + {\cal
    I}({c_{u_j}})\right]$ will lead to a misalignment between the
Radion couplings and the fermion mass
matrix. 
After diagonalizing the fermion mass matrix, flavor violating
couplings of the u-type quarks and radion will be generated, with
straightforward extensions to the down quark and lepton
sectors. We study these in Sec. \ref{sec:FCNC}, after describing the
flavor structure of the model with four generations.

%%%%%%%%%%%%%%%%%%%%%%%%%%%%%%%%%%%%%%%%%%%%%%%%%%%%%%%%%%%%%%%%%%%%%%%%%%%%%%%%
%%%%%%%%%%%%%%%%%%%%%%%%%%%%%%%%%%%%%%%%%%%%%%%%%%%%%%%%%%%%%%%%%%%%%%%%%%%%%%%%
\section{Flavor Structure with four families}
\label{sec:4generations} 

Warped scenarios have been used before to study an extension of the SM
with four generations \cite{Burdman:2009ih}. Our emphasis is here on
flavor-changing interactions between the radion and the fermions. The fermion
mass matrices are 
\bea
{\bf m_u} &=& F_Q\ Y_u\ F_u,\\\
{\bf m_d} &=& F_Q\ Y_d\ F_d,
\eea
where $F_Q$, $F_u$ and $F_d$ are $4\times4$ diagonal matrices whose
entries are given by the values at the IR brane of the corresponding zero-mode
wave functions:
\bea
F_Q\!= \!{\rm Diag}\! \left(
f_{Q_1} , 
f_{Q_2},
f_{Q_3},
f_{Q_4}\right),\     
F_u\!=\!{\rm Diag}\! \left(
f_{u_1} 
,f_{u_2}   
,f_{u_3},
f_{u_4} \right),\     
F_d\!=\!{\rm Diag}\! \left(
f_{d_1},f_{d_2},
f_{d_3},
f_{d_4} \right) \ \ \
\eea
where $f(c)$ is given by
\begin{equation}
\label{fc}
f(c) = \sqrt{\frac{1-2c}{1-(R/R^\prime)^{1-2c}}}~ .
\end{equation}
The matrices $Y_u$ and $Y_d$ are the 5-dimensional Yukawa
couplings, i.e. general $4\times4$ complex matrices. Because most of
the entries in the diagonal matrices $F_q$ are ``naturally''
hierarchical (for UV-localized fermions), the physical fermion mass
matrices $m_u$ and $m_d$ will 
inherit their hierarchical structure independently of the nature of
the true 5D Yukawa couplings $Y_u$ and $Y_d$, which can therefore contain all
of its entries with similar size (of order 1) and have no definite
flavor structure. This is the main idea behind scenarios of so-called
``flavor anarchy'', which we will consider also here, but applied to a
four-family scenario.

To diagonalize the mass matrices we use 
\bea
\hspace{4cm}  U_{Q_u}\ {\bf  m_u}\ W_u^\dagger &=& {\bf m_u^{diag}},
\label{UW}\\ 
\hspace{4cm}  U_{Q_d}\ {\bf  m_d}\ W_d^\dagger &=& {\bf m_d^{diag}}.  
\eea

One can in fact obtain a relatively simple formulation of the rotation
matrices $U_{Q_u}$, $U_{Q_d}$, $W_u$ and $W_{u}$ by expanding their
entries in powers of ratios $f_i/f_j$, where $i<j$ and with $i=1,2$
and $j=1,2,3,4$.
Keeping only the leading terms, we obtain for $U_{Q_u},~U_{Q_d}$ (see
\cite{Casagrande:2008hr} for the
  three family case):
\bea
U_{Q_u} &=& \left(
\begin{tabular}{cccc}
$1\ $ & $\vphantom{\frac{\int_\int}{\int_{\int_\int}}} \displaystyle
  \frac{[Y_u]_{{}_{21}}}{[Y_u]_{{}_{11}}}\ \frac{f_{Q_1}}{f_{Q_2}}\ \ $
&  ${\cal U}^{Q_u}_{13}\ \ $&  ${\cal U}^{Q_u}_{14}\ \ $
\\ 
$\vphantom{\frac{\int_\int}{\int_{\int_\int}}} -\displaystyle
  \frac{[Y_u]^{{}^*}_{{}_{21}}}{[Y_u]^{{}^*}_{{}_{11}}}\
\frac{f_{Q_1}}{f_{Q_2}}\ \ $
  & \ 1 & 
 ${\cal U}^{Q_u}_{23}\ \ $&  ${\cal U}^{Q_u}_{24}\ \ $
\\
$\ \ \vphantom{\frac{\int_\int}{\int_{\int_\int}}}
  \displaystyle\frac{[Y_u]^{{}^*}_{{}_{31}}}{[Y_u]^{{}^*}_{{}_{11}}}\
\frac{f_{Q_1}}{f_{Q_3}}\ $ & 
  $\vphantom{\frac{\int_\int}{\int_{\int_\int}}}
-\displaystyle\frac{[Y_u]^{{}^*}_{{}_{11,32}}}{[Y_u]^{{}^*}_{{}_{11,22}}}\    
\frac{f_{Q_2}}{f_{Q_3}}\ \ $   & $\cos \theta_{Q_u}$ & $\sin \theta_{Q_u}$\\  
$\vphantom{\frac{\int_\int}{\int_{\int_\int}}}
-  \displaystyle\frac{[Y_u]^{{}^*}_{{}_{41}}}{[Y_u]^{{}^*}_{{}_{11}}}\
\frac{f_{Q_1}}{f_{Q_4}}\ $  &   
 $\ \ \vphantom{\frac{\int_\int}{\int_{\int_\int}}}
  \displaystyle\frac{[Y_u]^{{}^*}_{{}_{11,42}}}{[Y_u]^{{}^*}_{{}_{11,22}}}\   
\frac{f_{Q_2}}{f_{Q_4}}\ \ $
  & $-\sin^*\theta_{Q_u}$& $\cos^*\theta_{Q_u}$\\
\end{tabular} \right),\label{UQ}
\eea 
and similarly for $Q_u \to Q_d$; 
and for the right-handed quark mixing matrix $W_{u}$:
\bea
W_{u} &=& \left(
\begin{tabular}{cccc}
$1\ $ &  $  \displaystyle
 \frac{[Y_u]^*_{{}_{12}}}{[Y_u]^*_{{}_{11}}}\ \frac{f_{u_1}}{f_{u_2}}\ \ $
& ${\cal W}^{u}_{13}\ \ $&  ${\cal W}^{u}_{14}\ \ $\\ 
$ \ \ \vphantom{\frac{\int_\int}{\int_{\int_\int}}} \displaystyle
  -\frac{[Y_u]^{{}}_{{}_{12}}}{[Y_u]^{{}}_{{}_{11}}}\ \frac{f_{u_1}}{f_{u_2}}\ \
$
  & \ 1 &${\cal W}^{u}_{23}\ \ $&  ${\cal W}^{u}_{24}\ \ $
  \\
$ \ \ \vphantom{\frac{\int_\int}{\int_{\int_\int}}}
  \displaystyle\frac{[Y_u]^{{}}_{{}_{13}}}{[Y_u]^{{}}_{{}_{11}}}\
\frac{f_{u_1}}{f_{u_3}}\ $ & 
  $\vphantom{\frac{\int_\int}{\int_{\int_\int}}}
-\displaystyle\frac{[Y_u]^{{}}_{{}_{11,23}}}{[Y_u]^{{}}_{{}_{11,22}}}\    
\frac{f_{u_2}}{f_{u_3}}\ \ $   & $\cos^* \theta_{u}$ & $\sin^* \theta_{u}$\\  
$\vphantom{\frac{\int_\int}{\int_{\int_\int}}}
-  \displaystyle\frac{[Y_u]^{{}}_{{}_{14}}}{[Y_u]^{{}}_{{}_{11}}}\
\frac{f_{u_1}}{f_{u_4}}\ $  &   
 $\ \ \vphantom{\frac{\int_\int}{\int_{\int_\int}}}
  \displaystyle\frac{[Y_u]^{{}}_{{}_{11,24}}}{[Y_u]^{{}}_{{}_{11,22}}}\   
\frac{f_{u_2}}{f_{u_4}}\ \ $
  & $-\sin \theta_{u}$& $\cos\theta_{u}$\\
\end{tabular} \right),
\label{Wu}
\eea 
with similar expression for $u \to d$.

Using the definition of the fermion mixing matrix $V_{CKM}$
\bea
V_{CKM}=U_{Q_u}^\dagger U_{Q_d}, 
\eea
we can obtain obtain for example a simple expression for $V_{us}$ as 
\bea
V_{us}=\frac{f_{Q_1}}{f_{Q_2}}\left(
\frac{[Y_d]_{{}_{21}}}{[Y_d]_{{}_{11}}}-
\frac{[Y_u]_{{}_{21}}}{[Y_u]_{{}_{11}}}\right).  
\eea
Similar expression can be obtained for other CKM mixing angles, as shown
in \cite{Frank:2011rw}. Since the 5D Yukawa matrix elements are
expected to be all of order 1, the observed hierarchies among the
CKM elements can be explained by hierarchies among the $f_i$ parameters.

%%%%%%%%%%%%%%%%%%%%%%%%%%%%%%%%%%%%%%%%%%%%%%%%%%%%%%%%%%%%%%%%%%%%%%%%%%%%%%%%
%%%%%%%%%%%%%%%%%%%%%%%%%%%%%%%%%%%%%%%%%%%%%%%%%%%%%%%%%%%%%%%%%%%%%%%%%%%%%%%%

\section{Flavor-Changing Neutral Couplings of the Radion}
\label{sec:FCNC}

The couplings between bulk SM fermions and the radion were calculated
in~\cite{CHL} for the case of one generation. Including the flavor
structure and the possibility of a bulk Higgs, these couplings are the
same for four generations as in the three-generation case,   
presented in~\cite{Azatov:2008vm} and take the form of
Eq. (\ref{Radioncoupling}).
After diagonalization of the fermion mass matrix, flavor violating
couplings will be generated. One can see this explicitly by performing
the bi-unitary rotation leading to the fermion mass basis, and writing
the radion couplings to fermions in that basis (in matrix form):
\bea
\hspace{-2cm}&&-\frac{\phi (x)}{\Lambda_\phi} {\bf \bar{d}}_L^{phys} 
\left[U^\dagger_{q_L} \hat{{\cal I}_q}U_{q_L}\ {\hat{m}_{d}^{diag}} +
  {\hat{m}_{d}^{diag}}\ W^\dagger_{d_R} \hat{\cal I}_dW_{d_R}\right] {\bf
d}_R^{phys}.
\eea
Here, ${\bf d}^{phys}$ is the physical state and is now a 4-vector in
flavor space, given that we have introduced an extra fourth
generation. Also we have defined $ \hat{{\cal I}_q}={\rm diag}[{\cal 
    I}({c_{q_i}})]$ and $ \hat{{\cal I}_d}={\rm diag}[{\cal I}({c_{d_i}})]$.
One observes that unless  the diagonal matrices $ \hat{{\cal I}_q}$ and
$\hat{{\cal
    I}_d}$ are both proportional to the unit matrix,\footnote{Note
  that this can easily be achieved if the bulk mass parameters, the 
  $c_i$'s, are all degenerate, but then the scenario cannot be used to
  produce/explain fermion hierarchies.}  there must be some
degree of flavor misalignment in the radion couplings.   
The extension to the up quark sector and charged leptons is
immediate.

%%%%%%%%%%%%%%%%%%%%%%%%%%%%%%%%%%%%%%%%%%%%%%%%%%%%%%%%%%%%%%%%%%%%%%%%%%%%%
\subsection{Radion FCNC's in Flavor Anarchy--Analytical Results}

We explicitly parametrize the radion couplings with fermions by showing the
mass dependence  as
\begin{equation}
\mathcal{L}_{q,FV} = - \frac{\tilde{a}^d_{ij}}{\Lambda_\phi} \sqrt{m_{d_{i}}
m_{d_{j}}}\ \phi \
\bar{d}^i_L d^j_R - \frac{\tilde{a}^u_{ij} }{\Lambda_\phi}\sqrt{m_{u_{i}}
m_{u_{j}}}\ \phi \
\bar{u}^i_L u^j_R+
\text{h.c.},\label{rfv}
\end{equation}
where $d^{i}, u^i$ are the quark mass eigenstates with masses $m_{d_{i}},
m_{u_{i}}$. 
Due to the simplicity of the flavor structure in the radion couplings, it is now
possible to give analytical expressions for these couplings, to leading
order in ratios of $f_i/f_j$. The general expressions are,  for $i<j$:   
\begin{eqnarray}
\label{ai<jexact}
{\tilde a}^d_{ij}&=& \sqrt{\frac{m_{d_j}}{m_{d_i}}} \sum_{k=1}^3 \left[ \left (
{ {\cal I}}( c_{q_k})-{ {\cal I}}( c_{q_4}) \right ) U^{Q_d* } _{ki}
U^{Q_d}_{kj}
\right ]\ \   +{\cal O} (\frac{m_{d_i}}{m_{d_j}}), \nonumber\\
{\tilde a}^u_{ij}&=& \sqrt{\frac{m_{u_j}}{m_{u_i}}} \sum_{k=1}^3 \left
[ \left ( { {\cal I}}( c_{q_k})-{ {\cal I}}( c_{q_4}) \right )
  U^{Q_u*} _{ki} U^{Q_u }_{kj} \right ]\ \  +{\cal O} (\frac{m_{u_i}}{m_{u_j}}), 
\end{eqnarray}
and for $i>j$:
\begin{eqnarray}
\label{ai>jexact}
{\tilde a}^d_{ij}&=& \sqrt{\frac{m_{d_i}}{m_{d_j}}} \sum_{k=1}^3 \left[ \left (
{ {\cal I}}( c_{d_k})-{ {\cal I}}( c_{d_4}) \right ) W^{d} _{ki} W^{d *}_{kj}
\right ]\ \   +{\cal O} (\frac{m_{d_j}}{m_{d_i}}),\nonumber\\
{\tilde a}^u_{ij}&=& \sqrt{\frac{m_{u_i}}{m_{u_j}}} \sum_{k=1}^3 \left
[ \left ( { {\cal I}}( c_{u_k})-{ {\cal I}}( c_{u_4}) \right ) W^{u}
  _{ki} W^{u *}_{kj} \right ]\ \ +{\cal O} (\frac{m_{u_j}}{m_{u_i}}).
\end{eqnarray}
Note that when $i<j$ the couplings are controlled by ``left-handed''
bulk masses ($c_{q}$) and mixings ($U^{Q}$), and when  $i>j$, the
couplings are controlled by ``right-handed''
bulk masses ($c_{u,d}$) and mixings ($W^{u,d}$).
The resulting $3\times 3$ substructure of these couplings, i.e. without the
fourth generation, match the results presented in \cite{Azatov:2008vm}.
%%%%%%%%%%%%%%%%%%%%%%%%%%%%%%%%%%%%%%%%%%%%%%%%%%%%%%%%%%%%%%%%%%%%%%%%%%%%

The expansion of the mixing angles in terms of ratios of $f$'s gives
$U^{Q_{(d,u)}}_{ij}  \sim f_{Q_i}/f_{Q_j},~W^{d}_{ij}  \sim 
f_{d_i}/f_{d_j},$ and $W^{u}_{ij}  \sim f_{u_i}/f_{u_j}$. With these,
we can obtain the parametric dependence of the radion couplings up to
corrections of order one.\footnote{See Appendix for details.} 

The diagonal terms are simply
\begin{eqnarray}
{\tilde a}^d_{ii}&\approx&{  {\cal I}}(c_{q_i})+{  {\cal I}}(c_{d_i}),~~~{\tilde
a}^u_{ii}\approx{  {\cal I}}(c_{q_i})+{  {\cal I}}(c_{u_i}).%\nonumber\\
%{\tilde a}^u_{ii}&=&{  {\cal I}}(c_{q_i})+{  {\cal I}}(c_{u_i}).
\end{eqnarray}
As the function ${\cal I}(c)$, defined in
  Eq.~(\ref{defI}), tends to $\ c\ $ for $c>1/2$, and approaches quickly the
value
  $1/2$ for $c<1/2$,   
 the diagonal terms in the down sector can be written as
\bea
{\tilde a}^d_{11} \approx (c_{q_1}+c_{d_1}),\ \  \ \ {\tilde
a}^d_{22}\approx(c_{q_2}+ c_{d_2}),\ \   \ \ 
{\tilde a}^d_{33} \approx (\frac{1}{2} + c_{d_3}),\ \    \ \ {\tilde
  a}^d_{44} \approx 1, \non
\eea
while the off-diagonal terms also get very simple expressions
\begin{eqnarray}
{\tilde a}_{12}^d \approx  \sqrt{\frac{m_{s}}{m_{d}}}\left(
    c_{q_1}- c_{q_2} \right)
\frac{f_{Q_1}}{f_{Q_2}} %{\cal O}(1) 
,&&\ \ 
{\tilde a}_{21}^d \approx
\sqrt{\frac{m_{s}}{m_{d}}}\left (  c_{d_1}- c_{d_2} \right )
\frac{f_{d_1}}{f_{d_2}} %{\cal O}(1)
,\nonumber\\ 
{\tilde a}_{13}^d \approx  \sqrt{\frac{m_{b}}{m_{d}}}  \left(  
c_{q_1} 
%-{ {\cal I}}( c_{q_2})+{ {\cal I}}( c_{q_3})
- \frac{1}{2} \right) \frac{f_{Q_1}}{f_{Q_3}}
%{\cal O}(1)
,&& \ \
{\tilde a}_{31}^d \approx  \sqrt{\frac{m_{b}}{m_{d}}}\left(
   c_{d_1} 
%-{ {\cal I}}( c_{d_2})+{ {\cal I}}( c_{d_3})
- c_{d_3} \right )\frac{f_{d_1}}{f_{d_3}}
%{\cal O}(1) ; 
,\nonumber\\
{\tilde a}_{23}^d \approx  \sqrt{\frac{m_{b}}{m_{s}}}\left(
   c_{q_2}- \frac{1}{2} \right)
\frac{f_{Q_2}}{f_{Q_3}} %{\cal O}(1) 
,&&\ \
{\tilde a}_{32}^d \approx
\sqrt{\frac{m_{b}}{m_{s}}}\left (c_{d_2}- c_{d_3} \right )
\frac{f_{d_2}}{f_{d_3}} %{\cal O}(1);
,\nonumber\\ 
{\tilde a}_{14}^d \approx  \sqrt{\frac{m_{b'}}{m_{d}}}
\left( c_{q_1} 
%-{ {\cal I}}( c_{q_2})+{ {\cal I}}( c_{q_3})
- \frac{1}{2}\right)\frac{f_{Q_1}}{f_{Q_4}} %{\cal O}(1)
,&& \ \
{\tilde a}_{41}^d \approx  \sqrt{\frac{m_{b'}}{m_{d}}}\left(
   c_{d_1} 
%-{ {\cal I}}( c_{d_2})+{ {\cal I}}(c_{d_3})
-\frac{1}{2} \right)\frac{f_{d_1}}{f_{d_4}}
%{\cal O}(1) ; 
,\nonumber\\
{\tilde a}_{24}^d \approx  \sqrt{\frac{m_{b'}}{m_{s}}}\left(
c_{q_2}-\frac{1}{2} \right)
\frac{f_{Q_2}}{f_{Q_4}} %{\cal O}(1) 
,&& \ \
{\tilde a}_{42}^d \approx
\sqrt{\frac{m_{b'}}{m_{s}}}\left (  c_{d_2}- \frac{1}{2} \right)
 \frac{f_{d_2}}{f_{d_4}} %{\cal O}(1);
,\nonumber\\ 
{\tilde a}_{34}^d  \approx \sqrt{\frac{m_{b'}}{m_{b}}}\left[
  { {\cal I}}( c_{q_3})- { {\cal I}}( c_{q_4}) \right]
\frac{f_{Q_3}}{f_{Q_4}} %{\cal O}(1) 
,&& \ \
{\tilde a}_{43}^d \approx
\sqrt{\frac{m_{b'}}{m_{s}}}\left(c_{d_3}-\frac{1}{2}\right)
 \frac{f_{d_3}}{f_{d_4}}. %{\cal O}(1).
\label{adparam}
\end{eqnarray}
\normalsize
Note that in the above we took $c_{q_3}\approx c_{q_4} \approx
c_{d_4}=1/2$ except in ${\tilde a}_{34}^d$ where the dominant term
comes from the (expected small) difference between $c_{q_4}$ and
$c_{q_3}$. 
\begin{figure}[t!]
\center
\vspace{-1cm}
\includegraphics[width=13cm,height=9cm]{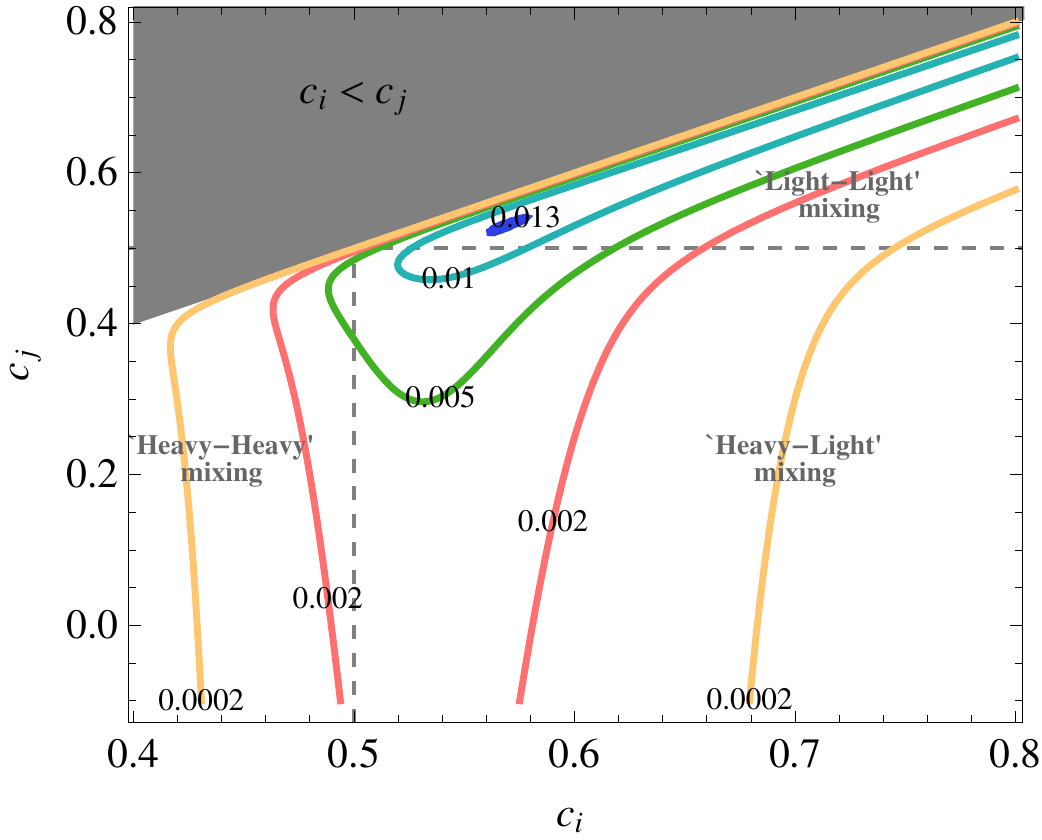}
\vspace{-.2cm}
\caption{Contours in the plane $(c_i,c_j)$ of the function
  $\hat{a}_{ij}= \left[  { {\cal I}}( c_{i})- { {\cal I}}(
    c_{j})  \right] \frac{f(c_i)}{f(c_j)}$, which sets the size of
  radion FCNC couplings with fermions. These are estimated to be
  ${\tilde a}_{ij}\simeq\sqrt{\frac{m_i}{m_j}}\hat{a}_{ij}$ and so from these
contours
  one can quickly estimate the size of these couplings by knowing the
  values of the bulk mass parameter $c_i$ of each fermion. }     
\label{aij}
\end{figure}
%%%%%%%%%%%%%%%%%%%%%%%%%%%%%%%%%%%%%%%%%%%%%%%%%%%%%%%%%%%%%%%%%%%%%

Similarly, in the up sector, we obtain
\bea
{\tilde a}^u_{11} \approx (c_{q_1}+c_{u_1}),\ \   \ \ {\tilde
a}^u_{22}\approx(c_{q_2}+ c_{u_2}), \ \  \ \
{\tilde a}^u_{33} \approx 1,\ \    \ \ {\tilde
  a}^u_{44} \approx 1, \non
\eea
and for the off-diagonal terms:
\begin{eqnarray}
{\tilde a}_{12}^u \approx  \sqrt{\frac{m_{c}}{m_{u}}}\left(
    c_{q_1}- c_{q_2} \right)
\frac{f_{Q_1}}{f_{Q_2}} %{\cal O}(1) 
,&&\ \ 
{\tilde a}_{21}^u \approx
\sqrt{\frac{m_{c}}{m_{u}}}\left (  c_{u_1}- c_{u_2} \right )
\frac{f_{u_1}}{f_{u_2}} %{\cal O}(1)
,\nonumber\\ 
{\tilde a}_{13}^u \approx  \sqrt{\frac{m_{t}}{m_{u}}}  \left(  
c_{q_1} 
%-{ {\cal I}}( c_{q_2})+{ {\cal I}}( c_{q_3})
- \frac{1}{2} \right) \frac{f_{Q_1}}{f_{Q_3}}
%{\cal O}(1)
,&& \ \
{\tilde a}_{31}^u \approx  \sqrt{\frac{m_{t}}{m_{u}}}\left(
   c_{u_1} 
%-{ {\cal I}}( c_{d_2})+{ {\cal I}}( c_{d_3})
- \frac{1}{2} \right )\frac{f_{d_1}}{f_{d_3}}
%{\cal O}(1) ; 
,\nonumber\\
{\tilde a}_{23}^u \approx  \sqrt{\frac{m_{t}}{m_{c}}}\left(
   c_{q_2}- \frac{1}{2} \right)
\frac{f_{Q_2}}{f_{Q_3}} %{\cal O}(1) 
,&&\ \
{\tilde a}_{32}^u \approx
\sqrt{\frac{m_{t}}{m_{c}}}\left (c_{u_2}- \frac{1}{2} \right )
\frac{f_{u_2}}{f_{u_3}} %{\cal O}(1);
,\nonumber\\ 
{\tilde a}_{14}^u \approx  \sqrt{\frac{m_{t'}}{m_{u}}}
\left( c_{q_1} 
%-{ {\cal I}}( c_{q_2})+{ {\cal I}}( c_{q_3})
- \frac{1}{2}\right)\frac{f_{Q_1}}{f_{Q_4}} %{\cal O}(1)
,&& \ \
{\tilde a}_{41}^u \approx  \sqrt{\frac{m_{t'}}{m_{u}}}\left(
   c_{u_1} 
%-{ {\cal I}}( c_{d_2})+{ {\cal I}}(c_{d_3})
-\frac{1}{2} \right)\frac{f_{u_1}}{f_{u_4}}
%{\cal O}(1) ; 
,\nonumber\\
{\tilde a}_{24}^u \approx  \sqrt{\frac{m_{t'}}{m_{c}}}\left(
c_{q_2}-\frac{1}{2} \right)
\frac{f_{Q_2}}{f_{Q_4}} %{\cal O}(1) 
,&& \ \
{\tilde a}_{42}^u \approx
\sqrt{\frac{m_{t'}}{m_{c}}}\left (  c_{u_2}- \frac{1}{2} \right)
 \frac{f_{u_2}}{f_{u_4}} %{\cal O}(1);
,\nonumber\\ 
{\tilde a}_{34}^u  \approx \sqrt{\frac{m_{t'}}{m_{t}}}\left[
  { {\cal I}}( c_{q_3})- { {\cal I}}( c_{q_4}) \right]
\frac{f_{Q_3}}{f_{Q_4}} %{\cal O}(1) 
,&& \ \
{\tilde a}_{43}^u \approx
\sqrt{\frac{m_{t'}}{m_{t}}}\left(  {\cal I}( c_{u_3})- { {\cal I}}( c_{u_4})
\right)
 \frac{f_{u_3}}{f_{u_4}}. %{\cal O}(1).
\label{auparam}
\end{eqnarray}
Here we assumed $c_{q_3}\approx c_{q_4} \approx c_{u_3} \approx
c_{u_4}=1/2$ except in ${\tilde a}_{34}^u$, ${\tilde a}_{43}^u$, for
the same reasons given for the down sector. This situation is very
different from the case of FCNC couplings of the Higgs boson
\cite{Frank:2011rw} where the couplings $a_{34},\,a_{43}$ are
large
due to significant misalignment in the 3-4 family.  
%%%%%%%%%%%%%%%%%%%%%%%%%%%%%%%%%%%%%%%%%%%%%%%%%%%%%%%%%%%%%%%%%%%%%%%%%%
%%%%%%%%%%%%%%%%%%%%%%%%%%%%%%%%%%%%%%%%%%%%%%%%%%%%%%%%%%%%%%%%%%%%%%%%%%
It is clear from the expressions for  ${\tilde a}_{ij}^d$,   ${\tilde
  a}_{ij}^u$ that the flavor changing couplings of the radion are of
the simple form $ \sqrt{\frac{m_j}{m_i}} \left[{\cal I}(c_{i})-{\cal
    I}(c_{j})\right] {\displaystyle \frac{f_i}{f_j}}$. We explore typical values
of
this function as contours in a $c_i, c_j$ plane, and determine the
localization coefficients  for which this function is maximal. In
Fig. \ref{aij}, we show contours of the  in the plane of the  ${\tilde
  a}_{ij}$ as a function of  two bulk mass parameters, $(c_i,c_j)$ for
$c_i<c_j$. The light-light regions correspond to mixing among the
first two families, and $b_R$. Corresponding to localization of light
quarks these are maximal  and the FCNC couplings of the radion
${\tilde a}_{ij}$ can reach a maximum of $0.013$. The heavy-light
mixing correspond to the fourth family mixing, or third family
doublet, or $t_R$ mixing, with the light two families and $b_R$. These
mixings can reach $0.01$, although they are more likely to be in the
$(0.002-0.005)$ region. Finally the heavy-heavy mixing (among fourth
families, $(t~b)_L$ and $u_R$) can reach $0.02$ as $c_{q_3}, c_{u_3}$
deviate from $1/2$. The results of the analytic calculations agree
with our numerical scan presented in the next subsection.

%%%%%%%%%%%%%%%%%%%%%%%%%%%%%%%%%%%%%%%%%%%%%%%%%%%%%%%%%%%%%%%%%%%%%%%%%%%%%
\subsection{Radion FCNC's in Flavor Anarchy--Numerical Results}

We complement our analytical consideration by performing a numerical
scan over the parameter space. We proceed as follows. We generate
random complex entries for $Y_u$ and $Y_d$, then obtain values for
$f_{u_i}, f_{d_i}$ and $f_{Q_i}$ in the same way as for the Higgs FCNC
couplings \cite{Frank:2011rw}, in matrix form. Using $f(c)$ from Eq. (\ref{fc}),
we solve for the
coefficients $c_i$. We then use the expression for ${\cal I}(c_i)$ to
calculate mass matrices ${\hat{\bf m_u}}, {\hat {\bf m_d}}$, then obtain the
eigenvalues, and the matrices $W_{u,d}$ and $U_{Q_{u,d}}$. We then
have all the ingredients to calculate the fermion-radion couplings.  
>From the scan in parameter space, we find the $\tilde{a}_{ij}^{d}$,
$\tilde{a}_{ij}^{u}$ as follows
\begin{equation}
\label{adnumerical}
 \tilde{a}_{ij}^{d}\sim 
\left( 
\begin{array}{cccc}
1.295-1.315 &~~~0.017-0.039 &~~~ 0.010-0.025 &~~~ 0.089-0.290\\
0.013-0.034 &~~~1.215-1.231 &~~~ 0.006-0.016 &~~~ 0.065-0.179\\
0.080-0.201 &~~~ 0.016-0.050 &~~~ 1.129-1.151 &~~~ 0.0002-0.001\\
0.024-0.076 &~~~ 0.018-0.049 &~~~ 0.004-0.012 & ~~~1.000-1.001
\end{array}
\right),
\end{equation}
\begin{equation}
\label{aunumerical}
\tilde{a}_{ij}^{u}\sim 
\left( \begin{array}{cccc}
1.294-1.320 & ~~~0.065-0.164 &~~~ 0.081-0.212 & ~~~0.094-0.268\\
0.022-0.055 &~~~ 1.135-1.158 &~~~ 0.019-0.047 &~~~ 0.019-0.053\\
0.030-0.098 & ~~~0.042-0.103 & ~~~1.002-1.016 &~~~ 0.0003-0.002\\
0.023-0.078 &~~~ 0.030-0.075 &~~~ 0.001-0.005 & ~~~1.000-1.002      
\end{array}
\right).
\end{equation}
The above ranges show the 50\% quantile of acceptable points, which
means that 25\% of points found predict lower ${\tilde a}^d_{ij},
\,{\tilde a}^d_{ij}$ values and 25\% of points predict higher values
than those shown in the matrices. The results of the scan are
consistent with the values obtained through analytical
considerations and from the values estimated using Fig. \ref{aij},
once typical sizes of the bulk masses are associated to the
appropriate fermions.

%%%%%%%%%%%%%%%%%%%%%%%%%%%%%%%%%%%%%%%%%%%%%%%%%%%%%%%%%%%%%%%%%%%%%%%%%%%%%
\subsection{Radion FCNC's in Flavor Anarchy--Leptons}
We proceed in a similar fashion to calculate the FCNC couplings of the
radion with the leptons. Assuming the neutrinos to be Dirac-type, we
parametrize the couplings as  
\begin{equation}
\mathcal{L}_{l,FV} = - \frac{\tilde{a}^l_{ij}}{\Lambda_\phi} \sqrt{m_{l_{i}}
m_{l{j}}}\ \phi \
\bar{l}^i_L l^j_R - \frac{\tilde{a}^\nu_{ij} }{\Lambda_\phi}\sqrt{m_{\nu_{i}}
m_{\nu_{j}}}\ \phi \
\bar{\nu}^i_L \nu^j_R+
\text{h.c.}\label{rfv}
\end{equation}
The couplings of the charged leptons will resemble those of the
down-type quarks, the only difference being that $c_{L_3} \neq
\frac12$. The coefficients $c_{L_i},~i=1,2,3$ are very close and can
be large, while $c_{L_4}=\frac12$.  The matrix $\hat{{\cal I}_L}={\rm
  diag}[{\cal  
    I}({c_{L_j}})],~j=1,2, 3,4$ in Eq. \ref{Radioncoupling} can be
written as a diagonal matrix plus a non-diagonal one, with entries
${\rm diag} (0,0,0, \Delta c)$, where $\Delta c=c_{L_4}-c_{L_i}$ can
be large.

As the neutrinos are massless, the only FCNC non-zero couplings
involve the fourth family, that is ${\tilde a}^{\nu}_{ij} \ne 0$ only
if either $i=4$ and/or $j=4$. While couplings with quark are restricted
by the CKM matrix ( its $3\times 3$ substructure), the lepton mixing
matrix $U^{PMNS}$ is not as well known, and thus
restrictions on the $U^{PMNS}_{i4},~ U^{PMNS}_{4j}$ are even less
established. We assume that the left-handed matrix $U^{L}$ is 
hierarchical, thus almost diagonal and $U^{\nu}$ nonhierarchical, and
almost the same as $U^{PMNS}$.  

This would imply the same type of parametric dependence as in the quark sector:
\begin{eqnarray}
{\tilde a}^\nu_{14}& =& \sqrt{\frac{m_{\nu_4}}{m_{\nu_1}}}
\left[ c_{L_1}-{{\cal I}}( c_{L_4}) \right ]\frac{f_{L_1}}{f_{L_4}} {\cal
O}(1),\nonumber\\
{\tilde a}^\nu_{24}&= & \sqrt{\frac{m_{\nu_4}}{m_{\nu_2}}}
\left[ c_{L_2}
-{\cal I}( c_{L_4}) \right ]\frac{f_{L_2}}{f_{L_4}} {\cal O}(1), \nonumber\\
{\tilde a}^\nu_{34}&= & \sqrt{\frac{m_{\nu_4}}{m_{\nu_3}}}\left [
  c_{L_3}-{\cal I}( c_{L_4}) \right ]\frac{f_{L_3}}{f_{L_4}} {\cal
  O}(1), 
\label{alparam}
\end{eqnarray}
where the coefficients $c_{L_i}$ describe the localization of the
lepton doublets and can be large, and $f_{L_i}$ are the values of the
zero-mode wavefunctions for the doublet lepton $i$ at the IR
brane. For the right-handed neutrinos, $W^\nu$ is almost diagonal to
insure small neutrino masses. Thus ${\tilde a}_{4j},~ j \ne 4$ are
small and can be neglected. 
Here the dominant term could be ${\tilde a}^\nu_{34}$.  For $c_{L_i} ,
c_{\nu_i} > 1/2$, 
 the zero modes wavefunctions are localized towards the UV brane; if 
$c_{L_i} , c_{\nu_i}< 1/2$, they are localized towards the IR
 brane. However the size of ${\tilde a}^\nu_{34}$ is also determined
 by the mixing terms $U^L_{33}U^{L*}_{34}$, as given in Eq. (\ref{ai<jexact}).
Thus the mixing will be proportional to $f(c) \equiv
\sqrt{\frac{1-2c}{1-\epsilon^{1-2c}}}$ where the hierarchically small
parameter $\epsilon=R/ R'\approx 10^{-15}$ is generally
referred to as the warp factor. Thus, we must choose a value for
$c_{L_3}$ which maximizes the expression  ${\tilde a}^\nu_{34} \approx
\left [  {\cal I}(c_{L_3}- {\cal I}(c_{L_4})\right]
f(c_{L_3})$.  These values correspond to the region of Fig. \ref{aij} for the
light-heavy region.

%%%%%%%%%%%%%%%%%%%%%%%%%%%%%%%%%%%%%%%%%%%%%%%%%%%%%%%%%%%%%%%%%%%%%%%%%%%%%%%%
%%%%%%%%%%%%%%%%%%%%%%%%%%%%%%%%%%%%%%%%%%%%%%%%%%%%%%%%%%%%%%%%%%%%%%%%%%%%%%%%
\section{Phenomenology}
\label{sec:pheno}
%%%%%%%%%%%%%%%%%%%%%%%%%%%%%%%%%%%%%%%%%%

\subsection{Bounds on Radion-mediated FCNC couplings}

The off-diagonal Yukawa couplings induce FCNC in both quark and
lepton interactions, which 
affect low energy observables and also give possible signatures
at colliders. In this section, we discuss restrictions on radion 
flavor violation coming from tree-level processes $\Delta F = 2$, such as
$K-\bar{K}$, $B- \bar{B} $, $D-\bar{D} $ mixing. We use an effective
Lagrangian approach \cite{Buchmuller:1985jz,delAguila:2000aa} to isolate the
contributions. 
The $\Delta F = 2$ process are described by the 
Hamiltonian \cite{UTfit,BurasWeakHamiltonian}
\begin{eqnarray}
{\cal H}_{eff}^{\Delta F =2} = \sum_{a=1}^{5} C_a Q_a^{q_i q_j} +
\sum_{a=1}^3 \tilde{C}_a \tilde{Q}_a^{q_i q_j}, 
\end{eqnarray}
with
\begin{eqnarray}
Q_1^{q_i q_j} &=& \bar{q}^\alpha_{jL}\gamma_\mu
q_{iL}^\alpha\bar{q}^\beta_{jL}\gamma^\mu q^\beta_{iL}, \quad
Q_2^{q_i q_j} = \bar{q}^\alpha_{jR} q_{iL}^\alpha
\bar{q}^\beta_{jR} q_{iL}^\beta, \quad Q_3^{q_i q_j} =
\bar{q}^\alpha_{jR}q_{iL}^\beta \bar{q}_{jR}^\beta q_{iL}^\alpha
, \\ \nonumber Q_4^{q_i q_j} &=& \bar{q}^\alpha_{jR}q_{iL}^\alpha
\bar{q}_{jL}^\beta q_{iR}^\beta, 
\quad 
Q_5^{q_i q_j}=
\bar{q}^\alpha_{jR} q_{iL}^\beta \bar{q}^\beta_{jL} q_{iR}^\alpha ,
\end{eqnarray}
where $\alpha, \beta$ are color indices. The operators $\tilde{Q}_a$
are obtained from $Q_a$ by exchange $L \leftrightarrow R$. For
$K-\bar{K}$ , $B_d-\bar{B}_d $, $B_s-\bar{B}_s$, $D-\bar{D}$
mixing, $q_i q_j = s d$, $b d$, $ b s$ and $ u c$ respectively.

Exchange of the flavor-violating radions gives rise to additional
contributions to
$C_2$,
$\tilde{C}_2$ and $C_4$ operators.  These
 are given below, using the model-independent 
bounds on BSM contributions as in \cite{UTfit} to present coupled
constraints on $\tilde {a}_{ij}$ couplings and  the radion mass $m_\phi$.  

\begin{figure}[t]
\center
\vspace{-1cm}
\includegraphics[width=15cm,height=10cm]{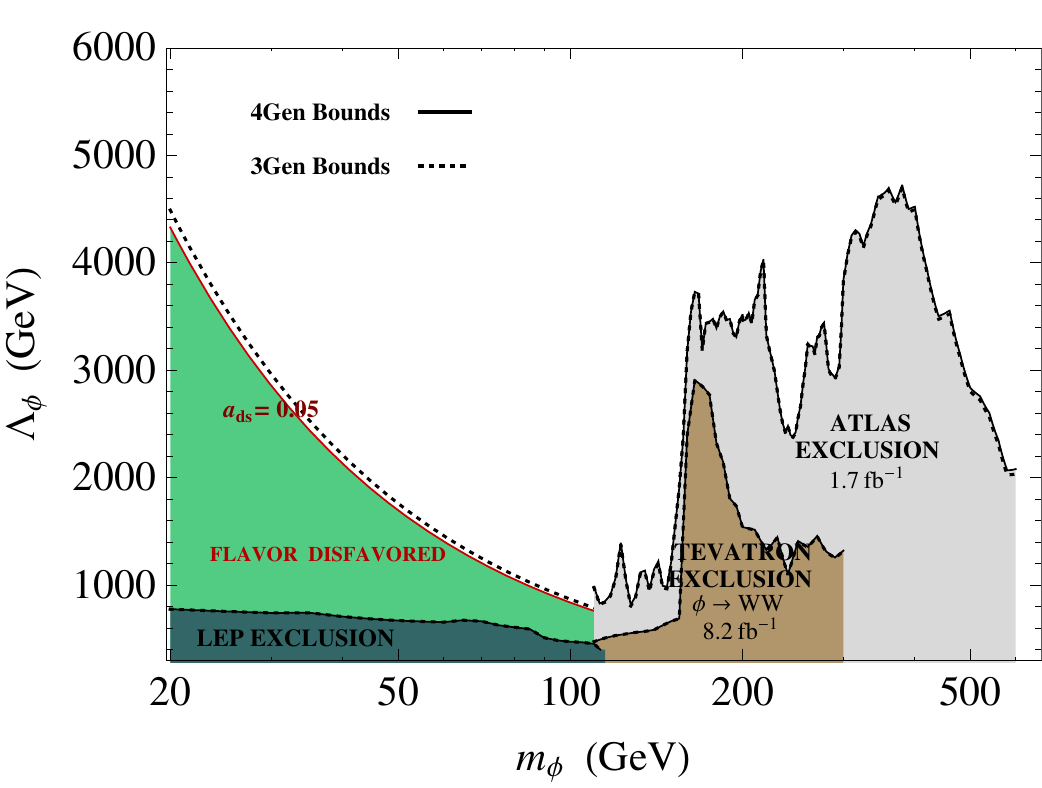}
\vspace{-.2cm}
\caption{Restrictions in the $m_\phi-\Lambda_\phi$ plane from collider
  exclusion limits and flavor constraints for $\epsilon_K$ (we have
  defined
  $a_{ds}=\sqrt{\textrm{Im}(\tilde{a}_{12}^{d*}\tilde{a}_{21}^d)}$).
One sees that for lighter radion ($m_\phi<160$ GeV) direct bounds are
quite weak and flavor physics provide stronger constraints (although less
robust). Heavier radions are mostly constrained by the ``golden mode''
$pp\to \phi\to ZZ$ and also $pp\to \phi\to WW$ at the LHC, while $p\bar p \to
\phi \to
WW$ is used at Tevatron.
 }      
\label{bounds}
\end{figure}

At the scale $m_{\phi}=60$ GeV, the limits on the $C_2$,
$\tilde{C}_2$ and $C_4$ operators are:
\begin{eqnarray}
 \textrm{ReC}_{K}^{2}&\leq& (\frac{1}{5.3\times
10^6 {\rm~GeV}})^2,\qquad\textrm{ReC}_{K}^{4}\leq (\frac{1}{9.1\times 10^6
{\rm~GeV}})^2,\nonumber\\
 \textrm{ImC}_{K}^{2}&\leq& (\frac{1}{9.5\times
10^7 {\rm~GeV}})^2,\qquad\textrm{ImC}_{K}^{4}\leq (\frac{1}{1.2\times 10^8
{\rm~GeV}})^2,\nonumber\\
|C_{D}^{2}|&\leq& (\frac{1}{1.8\times
10^6 {\rm~GeV}})^2,\qquad |C_{D}^{4}|\leq (\frac{1}{2.6\times 10^6
{\rm~GeV}})^2,\nonumber\\
|C_{B_{d}}^{2}|&\leq& (\frac{1}{8.7\times
10^5 {\rm~GeV}})^2,\qquad |C_{B_{d}}^{4}|\leq (\frac{1}{1.3\times 10^6
{\rm~GeV}})^2,\nonumber\\
|C_{B_{s}}^{2}|&\leq& (\frac{1}{1.0\times
10^5 {\rm~GeV}})^2,\qquad |C_{B_{s}}^{4}|\leq (\frac{1}{1.6\times 10^5
{\rm~GeV}})^2.
\end{eqnarray}
Using these bounds we obtain the constraints on radion flavor violating Yukawa
couplings (as compared to the $\tilde {a}_{ij}$ in the scan)
\footnotesize
\begin{eqnarray}
\Omega^{2}\textrm{Re}(\tilde{a}_{12}^{d*})^{2}&\leq& 2.6,\qquad 
\Omega^{2}\textrm{Re}(\tilde{a}^d_{21})^{2}\leq 2.6,\qquad 
\Omega^{2}\textrm{Re}(\tilde{a}_{12}^{*d}\tilde{a}^d_{21})\leq
0.90,\nonumber\\
\Omega^{2}\textrm{Im}(\tilde{a}_{12}^{d*})^{2}&\leq&
0.0082,\qquad 
\Omega^{2}\textrm{Im}(\tilde{a}_{21}^d)^{2}\leq 0.0082,
\qquad 
\Omega^{2}\textrm{Im}(\tilde{a}_{12}^{d*}\tilde{a}_{21}^d)\leq  0.0050,
\nonumber\\
\Omega^{2}|\tilde{a}_{13}^{u*}|^{2}&\leq& 3.2,\qquad
\Omega^{2}|\tilde{a}^u_{31}|^{2}\leq 3.2,\qquad
\Omega^{2}|\tilde{a}_{31}^{u*}\tilde{a}^u_{13}|\leq 1.4,\nonumber\\
\Omega^{2}|\tilde{a}_{13}^{d*}|^{2}&\leq& 1.9,\qquad
\Omega^{2}|\tilde{a}^d_{31}|^{2}\leq 1.9,\qquad
\Omega^{2}|\tilde{a}_{13}^{d*}\tilde{a}^d_{31}|\leq 0.87,\nonumber\\
\Omega^{2}|\tilde{a}_{32}^{d*}|^{2}&\leq& 6.5,\qquad
\Omega^{2}|\tilde{a}^d_{23}|^{2}\leq 6.5,\qquad
\Omega^{2}|\tilde{a}_{32}^{d*}\tilde{a}^d_{23}|\leq 2.8,
\end{eqnarray}
\normalsize
where $\Omega=\bigg(\frac{60~
{\rm GeV}}{m_{\phi}}\bigg)\bigg(\frac{2~
{\rm TeV}}{\Lambda_{\phi}}\bigg)$.
 Using our analytic results, the bounds
translate parametrically on restrictions on the bulk mass parameters
of the appropriate fermions. From the $\epsilon_K$ bounds
\begin{eqnarray}
\textrm{Im}(\tilde{a}_{12}^{d*}\tilde{a}_{21}^d) &&=-\frac{m_s}{m_d}(c_{q_1}-
c_{q_2})(c_{d_1}-c_{d_2}) \frac {f_{Q_1}f_{d_1}}{f_{Q_2}f_{d_2}}\textrm{Im}\left
(\frac{ [Y_d]_{21}^*[Y_d]_{12}^*}{([Y_d]_{11}^*)^2}\right)\nonumber \\
&&={\cal O}(1) (c_{q_1}-c_{q_2})(c_{d_1}-c_{d_2}),
\end{eqnarray}
where in the last expression we used the hierarchic nature of the
Yukawa couplings.\footnote{See Appendix for details.} This is a
remarkable result, as it relates the magnitude of $\epsilon_K$
directly to the bulk mass parameters (or the 
localization coefficients) of the $d,\,s$ quarks in the $U(1)_R$
singlet  and $SU(2)_L$ doublet representations. 
Similarly we can obtain appropriate expressions to obtain parametric
restrictions on the bulk mass parameters coming from $B^0-{\bar B}^0$
and $D^0-{\bar D}^0$ mixing: 
\begin{eqnarray}
|\tilde{a}_{13}^{d*}\tilde{a}_{31}^d| 
&=& {\cal O}(1) (c_{q_1}-\frac12)(c_{d_1}-c_{d_3}), 
 \nonumber\\
|\tilde{a}_{32}^{d*}\tilde{a}_{23}^d| &=&
{\cal O}(1) (c_{q_2}-\frac12)(c_{d_2}-c_{d_3}), \nonumber\\
|\tilde{a}_{13}^{u*}\tilde{a}_{31}^u| 
&=& {\cal O}(1) (c_{q_1}-\frac12)(c_{d_1}-\frac12). 
\end{eqnarray}
%%%%%%%%%%%%%%%%%%%%%%%%
One can see from the bounds, that unless the radion is very light
($m_\phi \sim 10$ GeV), the most significant constraints come from the
$\epsilon_K$ bounds, especially those on the coefficient $C_4$.
We thus use these bounds as the main flavor constraints on our model,
and present the restrictions in Fig. \ref{bounds} in the $m_\phi-\Lambda_\phi$
plane (for the typical value of ${\tilde a}^{d}_{12}\sim {\tilde a}^{d}_{21}\sim
0.05$).  The region below the $a_{ds}=0.05$ curve is thus named
``flavor disfavored'', since typical flavor anarchy parameter points
would produce too large contributions to $\epsilon_K$ in that region.
Note that we considered the scenarios with both 3 and 4 generations of
fermions, and the bounds are basically the same. The small difference
is due to the renormalization group running of operators, which is
slightly altered by the presence of extra fermion families.

In the same figure we present the most recent direct bounds on radion
phenomenology coming from collider data. Indeed one can easily use the
existing Higgs bounds to restrict regions in the
$m_\phi-\Lambda_\phi$ plane, since the search strategy for both the Higgs
and the radion are identical. This is due to the fact that the
couplings of the radion with matter particles are proportional
to the mass of the particles (just like the couplings of the Higgs).
The main difference is that the Higgs couplings are controled by the
electroweak scale $v$, whereas the radion couplings are controlled
(suppressed) by the much heavier scale $\Lambda_\phi$.

LEP bounds \cite{lephiggs} do apply for very light radion, although the
restrictions on $\Lambda_\phi$ are not too strong, and one sees that in
that region the generic flavor bounds are much stronger (although less
robust).

For heavier radion, the Tevatron and very recently the LHC do put
strong bounds in the allowed parameter region of our scenario. In both
experiments, the main production mechanism for the radion is via gluon
fusion but, unlike the Higgs, the other possible production mechanisms
such as vector boson fusion or associated W and Z production are
extremely suppressed. This is due to the special enhancement of the coupling of 
radion to gluons through the trace anomaly. The consequence of this fact is that
Higgs searches must be appropriately translated into radion bounds by
subtracting events coming from scalars produced via vector boson
fusion. One can do this roughly by adjusting the production cross
section of the Higgs in order to only obtain the gluon fusion cross
section. A better way of translating Higgs searches into radion is to
use fourth generation Higgs searches. This is because a Higgs with 4
generations will mainly be produced in gluon fusion (with almost no
other production channel) and so there will be no need of subtracting
events coming from other production mechanisms.

Another important issue when translating Higgs bounds is that the width of
a heavier Higgs ($m_h>200$GeV) starts to be relevant (i.e. becomes larger
than the experimental resolution). This means that more background events must
be
integrated in order to optimize signal events. But the radion width is
always going to remain much smaller than experimental resolution due
to its couplings being suppressed by $\Lambda_\phi$ (and not $v$ as in
the Higgs case). We must therefore adjust again the Higgs limits in order to
take this
fact into account, since much less background events should be kept in
a pure radion search \cite{GRW}. 

With all this in mind, we  translate Tevatron and LHC bounds from
Higgs searches and show the excluded regions in Figure
\ref{bounds}. From the Tevatron collider, we use the 
CDF and D0 combined search for a fourth generation Higgs, which allows
interesting bounds up to masses of $m_\phi=300$ GeV \cite{Benjamin:2011sv}. This
search
focuses on the Higgs decay into pairs of $W$ bosons and makes use of
an integrated luminosity of $8.2$ fb$^{-1}$. As for the LHC,
we use the recent results from the ATLAS
experiment \cite{atlascombinationsummer}, in
which they perform a combination of different channels, with
integrated luminosities up to $1.7$ fb$^{-1}$. As one can see, LHC
data from a single experiment outperforms the Tevatron and quite
interesting bounds can be set up to a mass of $m_\phi=600$ GeV.
We note that because the relative importance of different channels is
not exactly the same for Higgs and radion (specially the branching of
the $\phi\to \gamma\gamma$ channel differs from $h\to \gamma\gamma$), in the
lower
mass region $m_h<160$ GeV we avoid the combination and use exclusively
the ATLAS limits from $h\to \gamma\gamma$ search.
Above that point the branchings of Higgs and radion into heavy vector
bosons are essentially the same, specially if we assume for the plot that the
fourth
generation of fermions (if it exists) is heavy enough, with masses
greater than $300$ GeV. 

Finally we note from this figure that radion phenomenology does not
really change due to the addition of a fourth family{\footnote{We
    assume here that the radion decays to fourth generation fermions
    (especially leptons and neutrinos) is negligible. For an
    alternative scenario, see next section.}}. This might seem 
surprising because the Higgs phenomenology is greatly affected by the
presence of a fourth family of fermions (specially fourth family quarks) due
to an
important enhancement in the Higgs production cross section. This does not
happen in the case of the radion, because its couplings with massless
gauge bosons are quite indifferent to the addition of new heavy degrees
of freedom. Even though the new added fields will produce new loop
contributions to $\phi\to gg$ or to $h\to\gamma\gamma$, their presence
will also alter the $\beta$ functions of the appropriate gauge groups,
which will affect the couplings of the radion to massless bosons through
the trace
anomaly. The new trace anomaly effects coming from a fourth family
will in fact cancel the previous loop contributions in the limit of
very heavy new states \cite{CHL}, and so basically the radion couplings to
photons and gluons remains the same, controlled only by the light degrees
of freedom of the theory \cite{Goldberger:2007zk}.

%%%%%%%%%%%%%%%%%%%%%
\section{Flavor changing radion decays in the 4 generation model}
\label{sec:decays}
\begin{figure}[t]
\center
\vspace{-1cm}
\includegraphics[width=15cm,height=10cm]{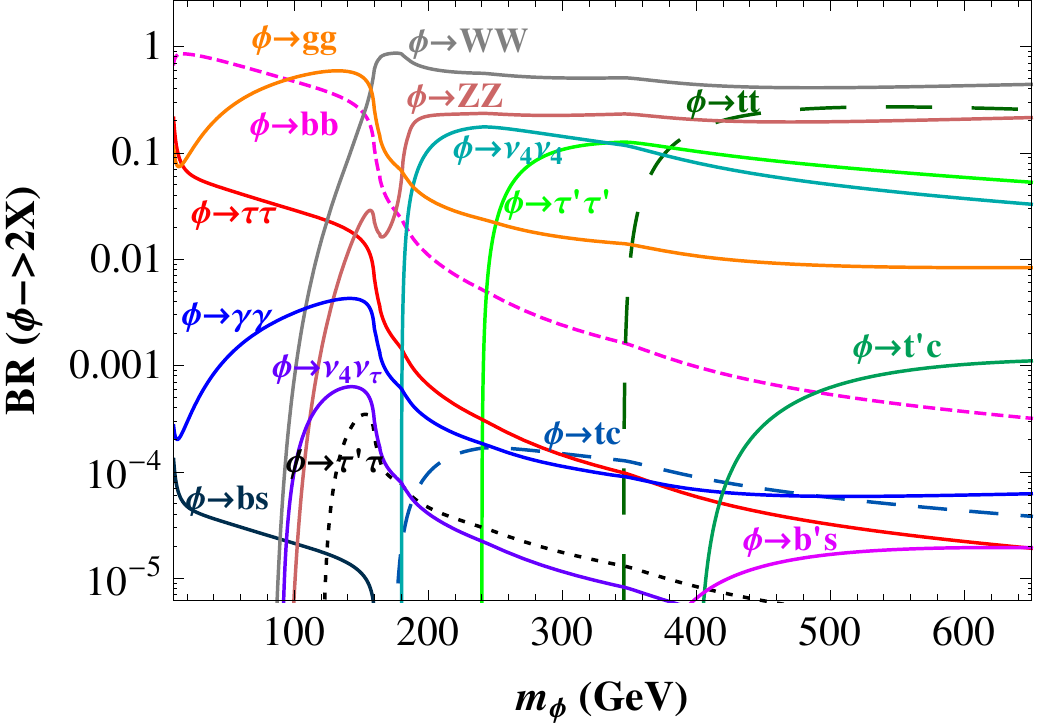}
\vspace{-.2cm}
\caption{Decay branching fractions of the radion in a warped scenario
  with four generations of fermions. 
The flavor anarchy setup (masses and mixings
  explained through fermion localization, with random 5D Yukawa
  couplings) predicts generic FV couplings of the radion, leading to a
  few new interesting decay channels such as $\phi \to bb'$ and $\phi \to
  \tau\tau'$. The masses chosen for this plot are $m_{b'}=350$ GeV,
  $m_{t'}=400$ GeV, $m_{\tau'}=120$ GeV and $m_{\nu_4}=90$ GeV, and the KK
scale is $(R')^{-1}= (\sqrt{6})1500$ GeV $ 
(\sim 3675$ GeV). }     
\label{radiondecays}
\end{figure}

%%%%%%%%%%%%%%%%%%%%%%%%%%%%%%%%%%%%%%%%%%%%%%%%%%%%%%%%%%%%%%%%%%%%%%%%%%%%%%%%
Radion couplings to fermions, massive and massless gauge bosons have
all been investigated before \cite{CHL,Azatov:2008vm}. Here we
investigate the changes in branching ratios due to the effect of a
fourth generation, and of flavor-changing interactions. We assume no
Higgs-radion mixing. We present our results in
Fig. \ref{radiondecays}. Note that we keep the radion mass to be above 
 $\sim 5-10$ GeV to avoid constraints from B-meson decays and
astrophysical data \cite{Davoudiasl:2009xz}.

%%%%%%%%%%%%%%%%%%%%%%%%%%%%%%%%%%%%%%%%%%%%%%%%%%%%%%

Depending on the masses of the fourth generation leptons and
neutrinos, FCNC decay channels ($\phi \to \tau \tau',~\nu_{\tau}
\nu_4$) could open for $m_\phi \ge100 $GeV.  
At higher radion masses, the $WW, ZZ$ and $t{\bar t}$ dominate. In
this region, the radion could be observed through the semi-leptonic
channel $\phi \to W_{lep}W_{had}$, and similarly $\phi \to t{\bar t} \to b
{\bar b}\,W_{had}W_{lep}$ (avoiding the fully hadronic channel which
suffers from large QCD dijet background), but the decays rate would be
comparable to that of a direct Higgs boson production.  

Finally, for light (Dirac) fourth-generation neutrinos or leptons,
near the present bounds, radion branching ratios to $\nu_4\nu_4$ and
$\tau' \tau'$ can be significant and compete with $ZZ$ and $WW$
decays, and thus significantly alter radion decay patterns for
$m_\phi>200$ GeV.

%%%%%%%%%%%%%%%%%%%%%%%%%%%%%%%%%%%%%%%%%%%%%%%%%%%%%%%%%%%%%%%%%%%%%%%%%%%%%%%%
%%%%%%%%%%%%%%%%%%%%%%%%%%%%%%%%%%%%%%%%%%%%%%%%%%%%%%%%%%%%%%%%%%%%%%%%%%%%%%%%

\section{Conclusions and Outlook}
\label{sec:conclusion}

In this work, we have investigated the phenomenology of the couplings,
especially the flavor-violating ones, of the radion to fermions in a
warped model with three and four generations where the fermions are
allowed to propagate in the bulk.   
We have shown how to obtain these couplings analytically, and
presented leading order expressions for them in a compact
form. Although the radion FCNC couplings have been analyzed before, 
some of the analytic expressions presented here are new. We also
explored the regions in which 
the couplings lie, and maximal values for these, as contour plots in a
plane defined by coefficients describing quark localization with
respect to the TeV brane. We are able to predict typical (and maximal)
values for the radion coupling to heavy-heavy, light-light, and
heavy-light quarks, and these results are confirmed by an extensive
numerical scan.  

Applying these to phenomenology of the radion, we calculated the
tree-level FCNC contributions to $K^0-{\bar K}^0$,  $\epsilon_K$,
$D^0-{\bar D}^0$ and $B^0-{\bar B}^0$ mixing, and the restrictions
imposed on the couplings. We obtain simple expressions relating quark
localization to these experimental values. The most stringent
constraints are from $\epsilon_K$, yielding a region of space in
$m_\phi-\Lambda_\phi$ parameter space  disfavored by flavor violation
consideration. We add to these the most recent constraints on Higgs
masses, from ATLAS and CMS, translating them into combined radion
mass-scale limits. Our figure shows that a large range around a light
radion mass-low scale ($\Lambda_{\phi} \sim 2$ TeV, $m_\phi \sim 60$ GeV)
survives. Our analysis also shows that, unlike the case of the Higgs
boson,  there are minute differences between radion mass-scale limits
in 3 and 4 generations, and thus these limits are quite independent of the
number of generations. This conclusion stands in the case where
the radion decays are not significantly influenced by decays into
fourth generation fermions (in particular to fourth generation
neutrinos or leptons, which have the lowest mass bounds).  
In a complete decay plot, we include all branching ratios of the
radion. Expected to be light, the radion decays primarily to $gg$ and
$b {\bar b}$ at low masses ($m_\phi \le 100$ GeV), while for heavier
radions ($m_\phi \approx 100$ GeV), FCNC decay channel such as $\nu_4
\nu_\tau$ (assuming Dirac neutrinos) and $\tau' \tau$ open, with
branching ratios of $10^{-3}$. These are the most promising FCNC
decays of the radion, barring the unlikely appearance of $\phi \to t'c$
at the high $m_{t'}=400$ GeV threshold. However, flavor-conserving
radion decays 
into fourth generation leptons and neutrinos can be large (for
$m_{\nu_4},\, m_{\tau'} \simeq 100$ GeV) and alter the
dominant decay modes for a heavier radion $r \to ZZ, WW$. These are
typical decay for a radion in a model with four generations and would 
provide a distinguishing signal for the model. If a heavy Higgs-like
state is discovered at the LHC with the usual ``golden mode'', $pp \to
h \to ZZ$, a width measurement could rule out a conventional Higgs
boson. A careful study for different and/or exotic decay channels of
that resonance might be the key to discover both a fourth generation
of fermions and a warped extra dimension.

%%%%%%%%%%%%%%%%%%%%%%%%%%%%%%%%%%%%%%%%%%%%%%%%%%%%%%%%%%%%%%%%%%%%%%%%%%%%%%%%
%%%%%%%%%%%%%%%%%%%%%%%%%%%%%%%%%%%%%%%%%%%%%%%%%%%%%%%%%%%%%%%%%%%%%%%%%%%%%%%%
\section{Acknowledgments}

M.T. would like to thank Nobuchika Okada for earlier collaboration and
discussions on translating Higgs and other experimental bounds into
radion bounds. M.T. would also like to thank Alex Azatov and Lijun Zhu
for discussions. We acknowledge NSERC of Canada for partial financial
support under grant number SAP105354.

\section{Appendix}

In the mass basis, the radion couplings with fermions are
\begin{equation}
\mathcal{L}_{q,FV} = - \frac{\tilde{a}^d_{ij}}{\Lambda_\phi} \sqrt{m_{d_{i}}
m_{d_{j}}}\ \phi \
\bar{d}^i_L d^j_R - \frac{\tilde{a}^u_{ij} }{\Lambda_\phi}\sqrt{m_{u_{i}}
m_{u_{j}}}\ \phi \
\bar{u}^i_L u^j_R+
\text{h.c.}\label{rfvA}
\end{equation}
where $d^{i}, u^i$ are the quark mass eigenstates with masses $m_{d_{i}},
m_{u_{i}}$. 
Due to the simplicity of the flavor structure in the radion couplings, it is now
possible to give analytical expressions for these couplings, to leading
order in ratios of $f_i/f_j$. In the down sector, the general expressions are, 
for $i<j$:   
\begin{eqnarray}
\label{ai<jexact}
{\tilde a}^d_{ij}&=& \sqrt{\frac{m_{d_j}}{m_{d_i}}} \sum_{k=1}^3 \left [ \left (
{ {\cal I}}( c_{q_k})-{ {\cal I}}( c_{q_4}) \right ) U^{Q_d* } _{ki}
U^{Q_d}_{kj} 
\right ]\ \   +{\cal O} (\frac{m_{d_i}}{m_{d_j}}), %% \nonumber\\
\label{aij1A}
\end{eqnarray}
and for $i>j$:
\begin{eqnarray}
\label{ai>jexact}
{\tilde a}^d_{ij}&=& \sqrt{\frac{m_{d_i}}{m_{d_j}}} \sum_{k=1}^3 \left
[ \left ( { {\cal I}}( c_{d_k})-{ {\cal I}}( c_{d_4}) \right ) W^{d}
  _{ki} W^{d *}_{kj} 
\right ]\ \   +{\cal O} (\frac{m_{d_j}}{m_{d_i}})%% \nonumber\\
.\label{aij2A}
\end{eqnarray}
where the function ${\cal I}(c)$ is defined by
\begin{eqnarray}
{\cal I}(c)= \left[\frac{(\frac{1}{2}-c)}{1-{(R/R')}^{1-2c}}+c\right] \approx
\Big\{ \begin{array}{c} c
  \, \,(\, c\, > \,1/2\,) \\ \frac{1}{2}\,\, (\, c \,<
  \,1/2\,) \end{array}.
\end{eqnarray}

Note that when $i<j$ the couplings are controlled by ``left-handed''
bulk masses ($c_{q}$) and mixings ($U^{Q}$), and when  $i>j$, the
couplings are controlled by ``right-handed''
bulk masses ($c_{u,d}$) and mixings ($W^{u,d}$).
%%%%%%%%%%%%%%%%%%%%%%%%%%%%%%%%%%%%%%%%%%%%%%%%%%%%%%%%%%%%%%%%%%%%

We can use the previous analytical expressions for these couplings 
to obtain surprisingly simple parametric dependences in terms of the 5D mass
parameters, up to leading order in ratios of the fermion masses 
$m_i/m_j$ and of wave function profiles $f_i/f_j$. The final results
were presented in the main text.  Here we show how to extract these
dependences carefully for a few terms.

We focus first on the couplings between the radion, the {\it down} quark
and the {\it bottom} quark. The couplings involved are $\tilde{a}^d_{13}$ and
$\tilde{a}^d_{31}$. From Eq.~(\ref{aij1A}) we can write explicitly 
\begin{eqnarray}
{\tilde a}^d_{13} &=& \sqrt{\frac{m_b}{m_d}} \Big[ \Big( c_{q_1}-
  {\cal I}( c_{q_4}) \Big) U^{Q_d *}_{11} U^{Q_d}_{13}+\Big( c_{q_2}-  {\cal I}(
c_{q_4}) \Big)
  U^{Q_d *}_{21}U^{Q_d}_{23} \non\\
&& \hspace{1cm} + \Big( {\cal I}( c_{q_3}) -  {\cal I}( c_{q_4}) \Big)
  U^{Q_d *}_{31}U^{Q_d}_{33}    \Big]\ +\ {\cal
  O}\left(\frac{m_d}{m_b}\right).\ \ \ \
\label{ad13exact}
\end{eqnarray}
We have assumed explicitly that ${\cal I}(c_{q_1})=c_{q_1}$ and ${\cal
  I}(c_{q_2})=c_{q_2}$ given that the left handed down and strange
quark are supposed to be UV localized. On the other hand
the left handed bottom quark and the left handed $b'$ quark are
TeV brane localized, and so their bulk mass parameters $c_{q_3}$ and
$c_{q_4}$ must be greater than $1/2$, and thus we have 
${\cal I}(c_{q_3})\simeq{\cal I}(c_{q_4})\simeq\frac12 $. 
We thus obtain the simpler expression (since ${\cal
  I}(c_{q_3})-{\cal I}(c_{q_4})\simeq 0$)
\bea
{\tilde a}^d_{13} &\simeq& \sqrt{\frac{m_b}{m_d}} \Big[ \Big( c_{q_1}-
  \frac12 \Big) U^{Q_d *}_{11} U^{Q_d}_{13}+\Big( c_{q_2}-  \frac12 \Big)
  U^{Q_d *}_{21}U^{Q_d}_{23}  \Big].\
\eea
Finally from the dependence of mixing angles with fermion profiles, 
i.e $U^{Q_{(d,u)}}_{ij}  \sim f_{Q_i}/f_{Q_j}$,  it is clear that the
two remaining terms are of the same order  $\sim
f_{Q_1}/f_{Q_3}$. Since $c_{q_1}$ is greater than $c_{q_2}$, we derive the
parametric result presented in the main text
\bea
{\tilde a}^d_{13} = \sqrt{\frac{m_b}{m_d}}  \Big( c_{q_1}-
  \frac12 \Big)   \frac{f_{Q_1}}{f_{Q_3}}\ {\cal O}(1).
\eea
A slightly different example is the calculation for ${\tilde a}^d_{31}$, which
we write first as
\bea
{\tilde a}_{31}^d
 &=& \sqrt{\frac{m_b}{m_d}} \Big[ \Big( c_{d_1}-
   c_{d_3} \Big) W^{d}_{11} W^{d*}_{13}+\Big( c_{d_2}-  c_{d_3} \Big)
  W^{d }_{21}W^{d*}_{23} \non\\
&& \hspace{1cm} + \Big( {\cal I}( c_{d_4}) -   c_{d_3} \Big)
  W^{d }_{41}W^{d*}_{43}    \Big]\ +\ {\cal
  O}\left(\frac{m_d}{m_b}\right).\label{ad31exact}
\eea
One difference now is that the right handed bottom quark is UV
localized and so we have ${\cal I}(c_{d_3})=c_{d_3} \neq
\frac12$. Also, now   $W^{d }_{41}W^{d*}_{43}\sim
\frac{f_{d_1}f_{d_3}}{f_{d_4}^2} \ll \frac{f_{d_1}}{f_{d_3}} \sim
W^{d}_{11} W^{d*}_{13}\sim  W^{d
}_{21}W^{d*}_{23}  $ and so the expression becomes
\bea
{\tilde a}_{31}^d
 &\simeq& \sqrt{\frac{m_b}{m_d}} \Big[ \Big( c_{d_1}-
   c_{d_3} \Big) W^{d}_{11} W^{d*}_{13}+\Big( c_{d_2}-  c_{d_3} \Big)
  W^{d }_{21}W^{d*}_{23} \Big]\non\\
 &=& \sqrt{\frac{m_b}{m_d}} ( c_{d_1}-
   c_{d_3})  \frac{f_{d_1}}{f_{d_3}}\  {\cal O}(1),
\eea
the last step simply using the fact that $(c_{d_1}-
c_{d_3})>(c_{d_2}-  c_{d_3})$, so that at least the parametric
dependence (up to order one corrections) is satisfied.

Finally we show the calculation for the coupling ${\tilde a}^d_{12}$,
between down and strange quarks, which can be obtained with
a slightly different choice of unitarity relations. 
We have
\begin{eqnarray}
{\tilde a}^d_{12} &=& \sqrt{\frac{m_s}{m_d}} \Big[ \Big( c_{q_1}-
   c_{q_2}\Big) U^{Q_d *}_{11} U^{Q_d}_{12}+ \Big(  {\cal I}( c_{q_3})- c_{q_2}
\Big)
  U^{Q_d *}_{31}U^{Q_d}_{32} \non\\
&& \hspace{1cm} + \Big( {\cal I}(c_{q_4})-c_{q_2} \Big)
  U^{Q_d *}_{41}U^{Q_d}_{42}    \Big]\ +\ {\cal
  O}\left(\frac{m_d}{m_s}\right).\ \ \ \ 
\label{ad12exact}
\end{eqnarray}
Taking into account hierarchies in the wavefunctions $f_i$, we can
directly neglect higher order terms and obtain
\begin{eqnarray}
{\tilde a}^d_{12} &=& \sqrt{\frac{m_s}{m_d}} \Big( c_{q_1}-
   c_{q_2}\Big)  \frac{f_{Q_1}}{f_{Q_2}}\ {\cal O}(1).
\end{eqnarray}
All other terms in the down and the up sector can be computed in the
same manner, and thus one can derive from precise analytical
expressions like those in Eqs.~(\ref{ad13exact}), (\ref{ad31exact})
and (\ref{ad12exact}), the much simpler approximative results
presented in the main text in Eqs.~(\ref{adparam}), (\ref{auparam})
and (\ref{alparam}).

%%%%%%%%%%%%%%%%%%%%%%%%%%%%%%%%%%%%%%%%%%%%%%%%%%%%%%%%%%%%%%%%%%%%%%%%%%%%%%%%
%%%%%%%%%%%%%%%%%%%%%%

%%%%%%%%%%%%%%%%%%%%%%%%%%%%%%%%%%%%%%%%%%%%%%%%%%%%%%%%%%%%%%%%%%%%%%%%%%%%%%%%
%%%%%%
%%%%%%%%%%%%%%%%%%%%%%%%%%%%%%%%%%%%%%%%%%%%%%%%%%%%%%%%%%%%%%%%%%%%%%%%%%%%%%%%
%%%%%%

\end{document}